\documentclass[epj]{svjour}
\usepackage{graphics}
\usepackage{amsmath}
\usepackage{amssymb}
\usepackage{manyfoot}%
\usepackage{booktabs}%
\usepackage{algorithm}%
\usepackage{algorithmicx}%
\usepackage{algpseudocode}%
\usepackage{listings}%
\usepackage{url}
\usepackage{hyperref}
\begin{document}

\title{Demonstration of a cryogenic, switchable electron source for low-temperature detector calibration} 
\author{
D.A.~Bennett\inst{1} \and
M.~Borghesi\inst{2,3} \and
P.~Campana\inst{2,3} \and
R.~Carobene\inst{2,3} \and
A.~Cattaneo\inst{2,3} \and
M.~De~Gerone\inst{4} \and
M.~Faverzani\inst{2,3} \and
L.~Ferrari~Barusso\inst{4,5} \and
E.~Ferri\inst{3} \and
S.~Gamba\inst{2,3}\thanks{email: s.gamba12@campus.unimib.it} \and
A.~Giachero\inst{2,3} \and
M.~Gobbo\inst{2,3} \and
D.~Labranca\inst{2,3} \and
J.~Mates\inst{1} \and
R.~Moretti\inst{2,3} \and
A.~Nucciotti\inst{2,3} \and
L.~Origo\inst{3} \and
D.R.~Schmidt\inst{1} \and
D.S.~Swetz\inst{1} \and
J.N.~Ullom\inst{1} 
}

\institute{
\inst{1} National Institute of Standards and Technology (NIST), 325 Broadway, Boulder, CO, USA \\
\inst{2} Dipartimento di Fisica “Giuseppe Occhialini”, Università degli Studi di Milano-Bicocca, Piazza della Scienza 3,
Milano, Italy \\
\inst{3} Istituto Nazionale di Fisica Nucleare (INFN), Sezione di Milano Bicocca, Piazza della Scienza 3,
Milano, Italy \\
\inst{4} Istituto Nazionale di Fisica Nucleare (INFN), Sezione di Genova, Via Dodecaneso 33, Genova, Italy \\
\inst{5} Dipartimento di Fisica, Università di Genova, Via Dodecaneso 33, Genova, Italy
}
\date{Received: date / Accepted: date}

\abstract{
    We present the realization of a prototype of a compact, switchable electron source operating at cryogenic temperatures, demonstrated at energies ranging from 100 eV to 300 eV, and conceptually extendable to an arbitrary energy range. The electrons are produced via photoelectric emission, induced by a LED illuminating two 400~nm-thick commercial aluminum layers inside a cryostat, and are subsequently accelerated by a voltage of up to 300~V. Detection is carried out with an array of transition-edge sensor (TES) microcalorimeters designed for X-ray detectors, where we successfully observe signals consistent with electrons produced at the source at rates of $\gtrsim$1~Hz and with an efficiency of $\gtrsim 10^{-14}\ e^-/\gamma$. 
    }
    

\maketitle
\flushbottom

\section{\label{sec:Intro}Introduction} 
We present the development of a novel compact, switchable electron source capable of operating at cryogenic temperatures. Achieving reproducible electron generation under these conditions marks a critical milestone, as it opens new opportunities for calibrating cryogenic detectors. 

Conventionally energy calibration, for energies ranging from few eV to tens of keV, is performed using the full absorption of X-rays with known energies, such as those from the decay of ${}^{55}$Fe. X-ray fluorescence can provide additional calibration lines at lower energies~\cite{COLLING1995408},\cite{PhysRevLett.127.061801},\cite{LEBLANC1996208},\cite{Abele2025}.
This method poses significant practical challenges in a cryogenic environment. A first important consideration is that the source must be compact and properly thermalized at temperatures of a few mK, which is not trivial to achieve. Moreover, calibration and data-taking are generally performed separately, since the radioactive source cannot be switched off. A solution to this issue was adopted in~\cite{SISTI2004125}, where a movable multi-line fluorescence source was mounted on a half-cylinder designed to fit inside a massive shield made of high-purity Roman lead, displaced with respect to the detector holders. This configuration was intended to shield the internal bremsstrahlung radiation accompanying the $^{55}$Fe electron-capture decay; however, such a solution introduces considerable mechanical complexity in a cryogenic environment. Moreover, the energy of the calibration line is upper-bounded, and may not coincide with the specific energy region of interest or cover the full dynamic range required for precise calibration. One example of an existing source is reported in~\cite{pepe_detection_2024-1}, where electrons are produced by field  emission from vertically aligned carbon nanotubes. Although this approach provides a compact and switchable source, its performance can be limited by emitter degradation and by the stability and uniformity of the field emission~\cite{Groning_Clergereaux_Nilsson_Ruffieux_Groning_Schlapbach_2002}. Moreover, in the proposed baseline implementation, the number of emitted electrons is proportional to the electron energy, preventing independent control of these two parameters.

We propose a novel calibration source that is switchable, cryogenic-compatible, and allows the electron yield to be adjusted independently of the electron energy. The proposed calibration source involves three main steps: generating electrons through photoelectric emission induced by a LED on a conversion material (photocathode) and accelerating them, amplifying the electrons using an electron multiplier, and finally inducing their interaction with a conversion material to produce characteristic X‑ray photons. This approach has already been adopted in~\cite{10.1117/12.855880},\cite{10.1117/12.924146}, although in those cases the setup is based on highly expensive industrial photocathodes and requires a tube evacuated to a very high vacuum. In contrast, our proposal makes use of a simple metallic photocathode.
In this study, we report the successful completion of the electron‑generation stage, demonstrating the production of electrons under cryogenic conditions using a dedicated experimental setup described in Section~\ref{sec:setup}. In Section~\ref{sec:tests}, we present preliminary tests validating the functionality of the setup; in Section~\ref{sec:caratterizzazione}, we focus on beam characterization, including its profile, rate, and efficiency; and in Section~\ref{sec:spectrum}, we attempt to investigate transition-edge sensor (TES) response to impinging low-energy electrons.

In particular, this technology could enable a novel fluorescent calibration source for the HOLMES experiment~\cite{Alpert_2015},\cite{Becker_2019}, \cite{alpert2025stringentboundelectronneutrino},\cite{ahrens2026holmes},\cite{bennett2025impactembedded163hoperformance}, which aims to directly measure the electron neutrino mass through calorimetric analysis of the $^{163}$Ho electron capture spectrum, which extends up to 2.863~keV~\cite{schweiger_penning-trap_2024b}.
The HOLMES experiment makes use of an array of ion-implanted TES microcalorimeters~\cite{bennett2025impactembedded163hoperformance},\cite{alpert_high-resolution_2019},\cite{Irwin2005}. Accurate calibration is essential because the recorded event amplitude is non-linear with respect to the deposited energy~\cite{Fowler_2018}. The intrinsic non-linearity of the TES response can distort the measured spectrum, biasing the determination of the neutrino mass, since each detector exhibits a distinct non-linear behavior. At present, the HOLMES measurement is dominated by statistical uncertainties, making this effect negligible. However, as higher statistics are achieved, the impact of non-linearity is expected to become significant. 

The current calibration strategy relies on self-calibration using the $^{163}$Ho spectrum itself. In a dedicated run, the HOLMES detectors are placed in front of an X-ray source, generated by irradiating sodium chloride and aluminum targets with a $^{55}$Fe source~\cite{ahrens2026holmes}. The resulting fluorescence lines are used to calibrate the detector response and to identify the $^{163}$Ho spectral peaks. During physics runs, the X-ray source is removed, and the previously identified peaks are used for the energy calibration. This introduces an unavoidable systematic uncertainty at the endpoint, since the highest calibration peak (M1) is $\sim 800$ eV below the endpoint, with no calibration peaks available at higher energies. An independent calibration line, generated by the novel source presented in this article, which produces electrons that stimulate the emission of X-rays through the atomic excitation in the target material, preferably at energies above the endpoint, would therefore be highly beneficial.

With reference to Figure \ref{fig:app}, an electron amplifier would be placed after the electron acceleration stage, while a conversion material would be placed before the HOLMES detectors. If the source operates in pulsed mode, such a calibration line would allow periodic monitoring of each detector pixel during data taking. This could be achieved by selecting only the short intervals containing the calibration pulses for the energy calibration, while preserving the remaining data for the neutrino-mass analysis.


Besides the source characterization, a significant outcome of this work is the demonstration of low-energy electrons detection ($E \lesssim 300$ eV) using TESs~\cite{Irwin2005},\cite{Ullom_2015}. While TESs are expected to be sensitive to such electrons~\cite{Patel_2021}, few studies investigated their response to very low-energy electrons, in the 300 to 2000~eV~\cite{patel_electron_2024-1} and 90 to 101~eV~\cite{pepe_detection_2024-1} ranges. We used the sensors specifically developed for the HOLMES experiment~\cite{Alpert_2015},\cite{Becker_2019},\cite{ahrens2026holmes},\cite{alpert2025stringentboundelectronneutrino}: while these are the intended target of the future calibration source, their successful operation here provides a general proof of the sensitivity of such detectors to impinging electrons.
\section{\label{sec:setup}Experimental setup}
\begin{figure}[h!]
\centering
\resizebox{0.8\textwidth}{!}{\includegraphics{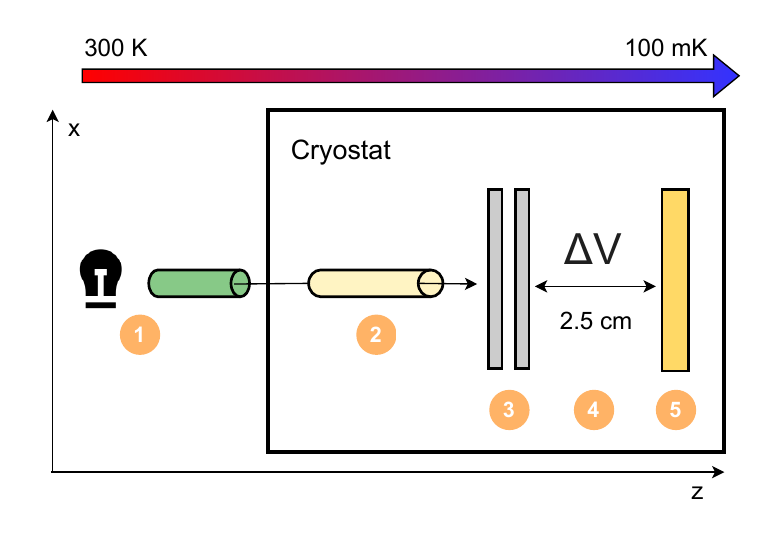}}
\caption{The experimental setup. The upper arrow indicates the temperature gradient from the exterior of the cryostat (300~K) to its interior (100~mK). Components are labeled as follows: (1) fiber-coupled LED, (2) vacuum fiber, (3) two layers of 400 nm Al, (4) high-voltage power supply connected to the Al foils and (5) calorimeter array (grounded to the cryostat). }
\label{fig:app}
\end{figure}

The core concept of this prototype setup (Figure~\ref{fig:app}) is to generate electrons through the photoelectric effect, by illuminating aluminum with photons to induce electron emission.
The setup operates in a dilution refrigerator (Oxford Instruments, Triton 200) with the mixing chamber stabilized at 60 mK. The TESs are biased to operate within their superconducting transition at roughly 100 mK.
Outside the cryostat, we use a M280F5 fiber-coupled LED with a core diameter of 400~$\mu$m (Figure~\ref{fig:app} (1)) and a wavelength of 280~nm (10~nm FWHM). Although the LED is rated for a maximum current of 500~mA-corresponding to an optical power of approximately 800~$\mu$W at room temperature-we restrict its operation to a bias current of approximately 60 to 180~mA in order to prevent excessive heating of the cryostat. 
The fiber used within the cryostat is an FVP100120140 vacuum-compatible fiber (Figure~\ref{fig:app} (2)), suitable for wavelengths ranging from 180~nm to 1150~nm, with an attenuation of approximately 0.3~dB/m and a length of 0.8~m. The fiber ends on two overlying 400~nm-thick Al foils (photocathode) (Figure~\ref{fig:app} (3)). The choice of this configuration is motivated in the following text. With this configuration, the photon flux incident on the photocathode is expected to be $\Phi \sim 10^{14} \ \gamma/\text{s}$ at $\sim$180 mA.
  
A simple metal was chosen as the photocathode despite its lower quantum efficiency compared to standard high-efficiency alkali-based materials because it can operate without requiring an ultra-high vacuum environment. Specifically, aluminum was selected among metals due to its relatively low work function (about 4~eV), compatibility with our LED wavelength, and commercial availability in thin films. The quantum efficiency of aluminum depends on photon energy, mean free path, and work function at a given temperature. However, since it is in practice highly sensitive to surface conditions and is well documented only for clean surfaces~\cite{CHEVALLAY1994146},\cite{RAHEMI201541}, the actual efficiency for commercial foils remains unknown. Furthermore, the physics of photoemission at ultralow temperatures involves complex transport mechanisms. While UV photons are absorbed within a short mean free path (a few nanometers), the subsequent electron attenuation length—which determines the escape probability—is unknown at low temperatures. Crucially, the impact of the superconducting transition ($T_c \approx 1.2$ K) on the photoemission process remains entirely unexplored in the literature. Finally, to address the issue of light leakage caused by potential microscopic pin-holes in commercial aluminum thin films, a double-layer configuration was adopted to ensure a better light-tight seal.

\begin{figure}[h!]
\centering
\resizebox{0.8\textwidth}{!}{\includegraphics{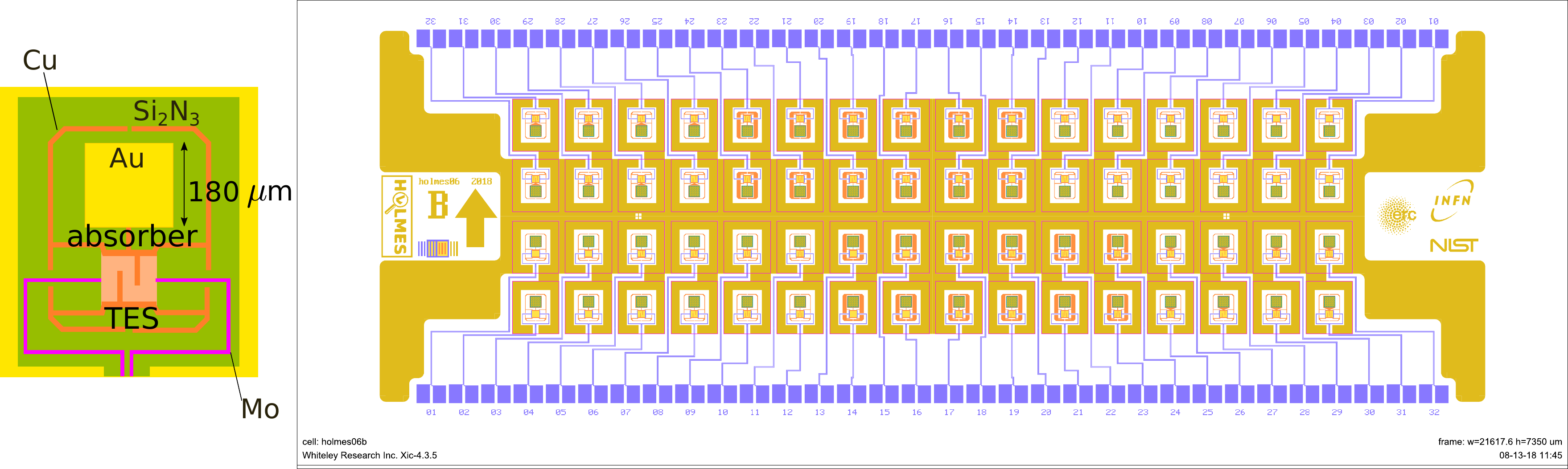}}
\caption{Left: Schematic, not to scale, representation of one of the TES microcalorimeter (channel 0). Right: The TES microcalorimeters 4$\times$16 sub-arrays. In the following, we refer to Channel 0 (1) as the thirteenth (twelfth) channel from the left in the second (first) row.}
\label{fig:arrayholmes}
\end{figure}

To detect electrons, we use one of the TES arrays fabricated for the HOLMES experiment~\cite{alpert2025stringentboundelectronneutrino}. These arrays are designed to operate in an energy range of $\lesssim$10 keV, with an energy resolution of $\sim$5 eV at 6 keV. Each detector in the array consists of a $180 \times 180 \times 2$ $\mu\text{m}^3$ gold absorber coupled to a Mo/Cu superconducting bilayer TES (Figure \ref{fig:arrayholmes}). Figure \ref{fig:arrayholmes} shows a 4$\times$16 detector array; however, the bottom two 2$\times$16 TES arrays on the chip were not wire-bonded and therefore were not readout. The resulting high-sensitivity microcalorimeters are readout using a microwave multiplexing scheme~\cite{alpert_high-resolution_2019} \cite{Becker_2019}, allowing us to recover the entire electrothermal response of the detectors. The array is placed 2.5~cm away from the aluminum foils in a detector box. Although the box features an opening above the array to permit electron access, the setup is wrapped in aluminum foil-heat-sunk at the refrigerator base temperature-to prevent thermal radiation from perturbing detector operation. No collimator was placed between the Al foils and the detectors, allowing electrons to strike areas of the chip outside the TES absorber. The TES box is maintained at the potential of the cryostat, which serves as the potential reference (Figure~\ref{fig:app} (5)). The aluminum foils are set to a negative voltage with respect to this reference (Figure~\ref{fig:app} (4)), thereby directing electrons towards the array for detection, with the exception of some tests in which the photocathode was set to a positive bias voltage. The voltage source is connected to the photocathode from outside the cryostat via a Nomex-textured Manganin twisted-pair cable routing from room temperature to the mixing chamber, thermally anchored at every cryostat stage. Since this wiring is not rated for high voltage, the polarization is limited to 300 V to prevent electrical breakdown.

\section{\label{sec:tests}Preliminary tests}
\begin{figure}[h!]
\centering
\resizebox{0.65\textwidth}{!}{\includegraphics{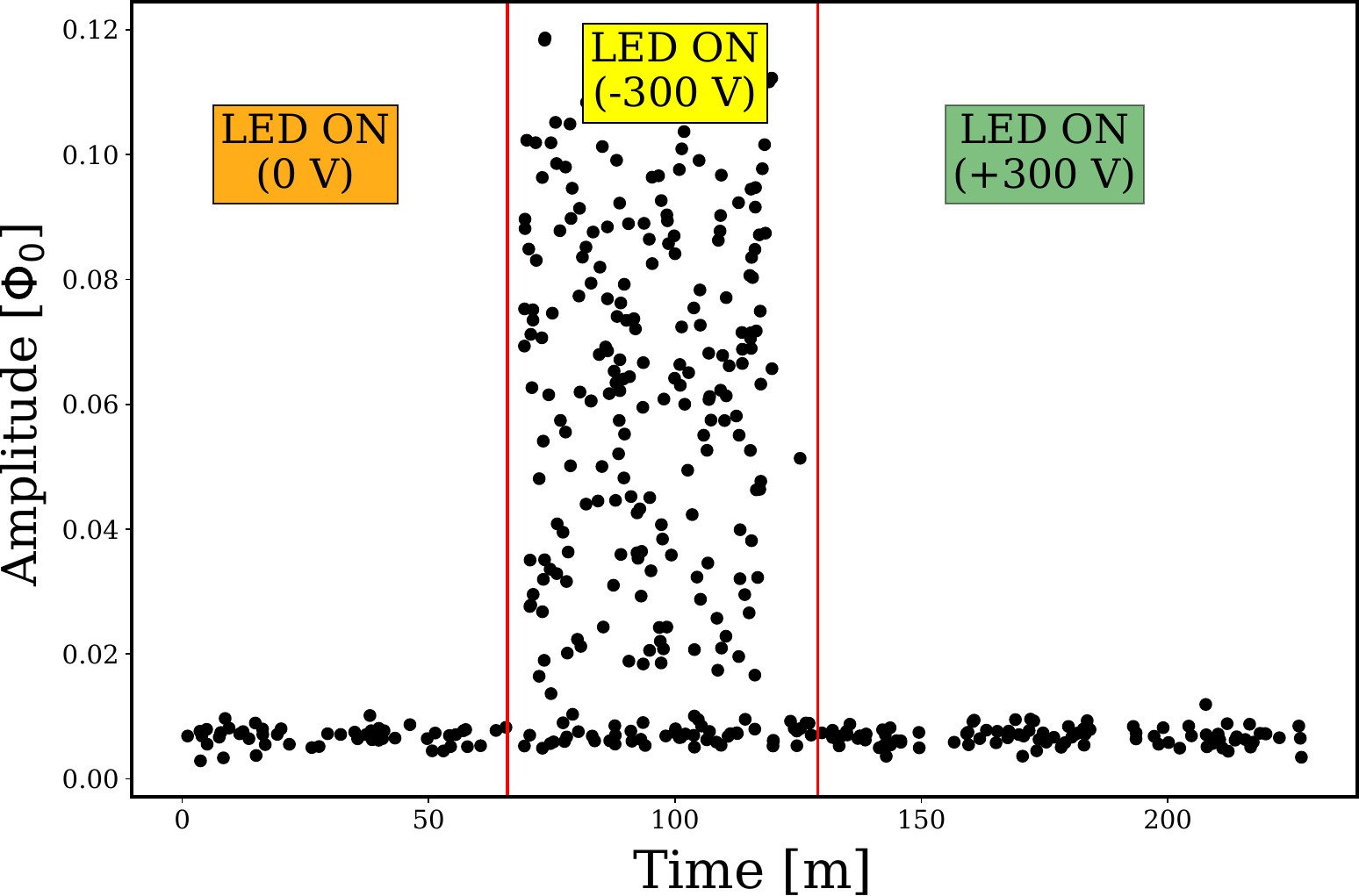}}
\caption{The peak-to-peak amplitude versus the arrival time in a test channel. Each data point represents a single event in a TES detector. By switching on the LED, three Al foils voltage values (0V, -300V, +300V) have been tested: a significant increase in counts was observed only with negative voltage. Noise contribution dominates at low amplitudes. }
\label{fig:analysis1}
\end{figure}

Preliminary measurements were conducted to validate the setup functionality and confirm that the observed TES signals are caused by photo-emitted electrons. To verify the negatively charged nature of the particles, the polarity of the bias voltage applied to the aluminum foils was reversed. The voltage difference between the Al foils and the TES box, which sits at 0 V, was varied in three configurations: 0 V, -300 V, and +300 V. The absence of signals in the 0 V and  +300 V configurations, as observed, supports the hypothesis that the detected particles are electrons. Figure \ref{fig:analysis1} shows the peak-to-peak amplitude\footnote{The peak-to-peak amplitude is measured as the difference between the maximum and minimum values of the pulse.} of continuously acquired TES signals as a function of time, divided into three time regions corresponding to the applied voltages. In this picture, $\Phi_0$ denotes the magnetic flux quantum ($\Phi_0 = h/2e$), the unit in which the signal measured by the rf-SQUIDs is expressed. The data indicate that a non-zero signal amplitude is observed only when a negative voltage is applied.

\begin{figure}[h!]
\centering
\resizebox{0.55\textwidth}{!}{\includegraphics{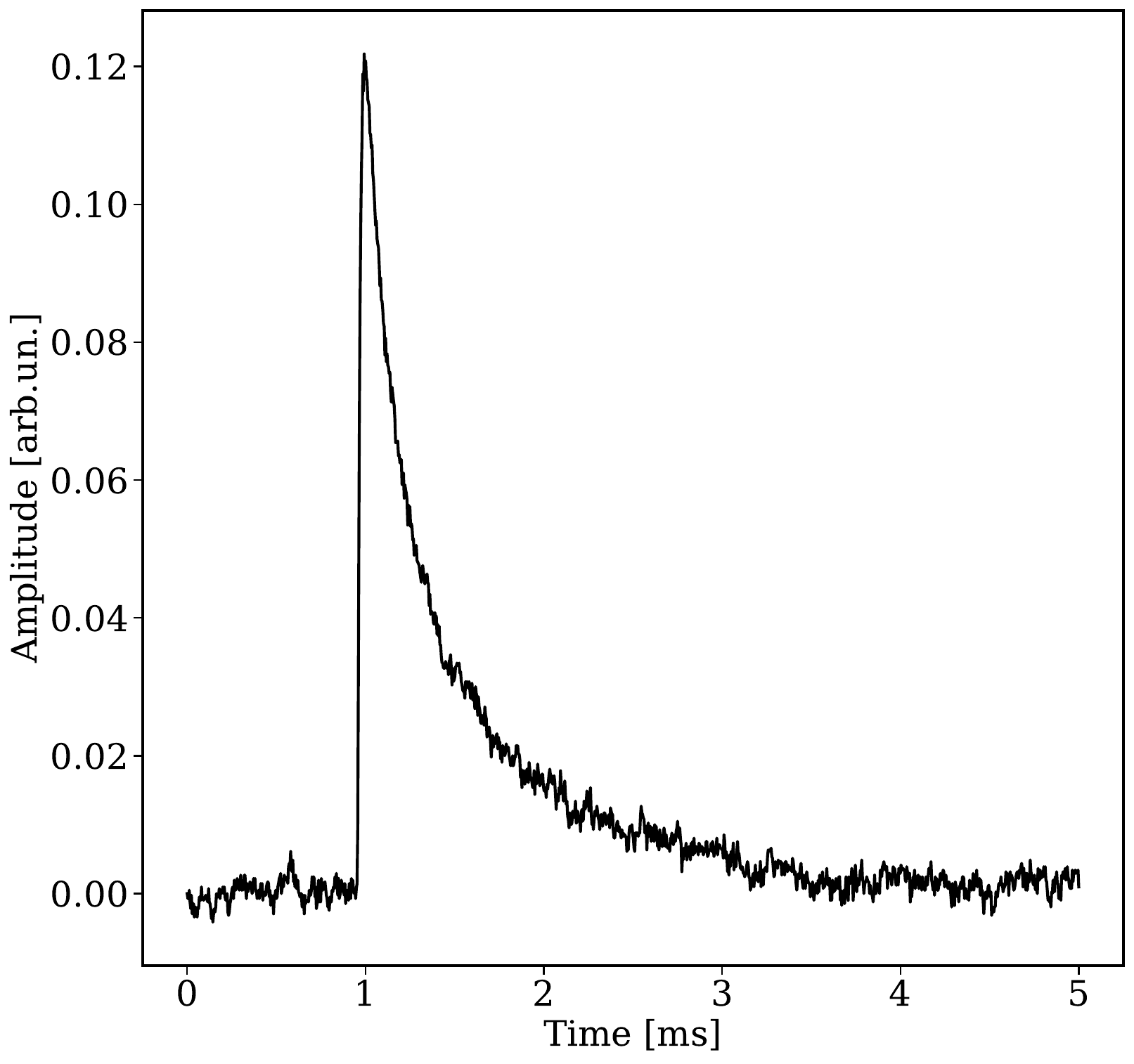}}
\caption{One of the acquired electron signals (channel 0) at -300 V.}
\label{fig:analysis99}
\end{figure}

In addition to this histogram, we also show one of the pulses acquired during the source characterization in Figure~\ref{fig:analysis99}, at a bias voltage of -300 V. This pulse exhibits the typical features observed in pulses acquired in the past with our detectors, namely a sharp rise time of about 10~$\mu$s and decay times of approximately 300~$\mu$s.

\begin{figure}[h!]
\centering
\resizebox{0.65\textwidth}{!}{\includegraphics{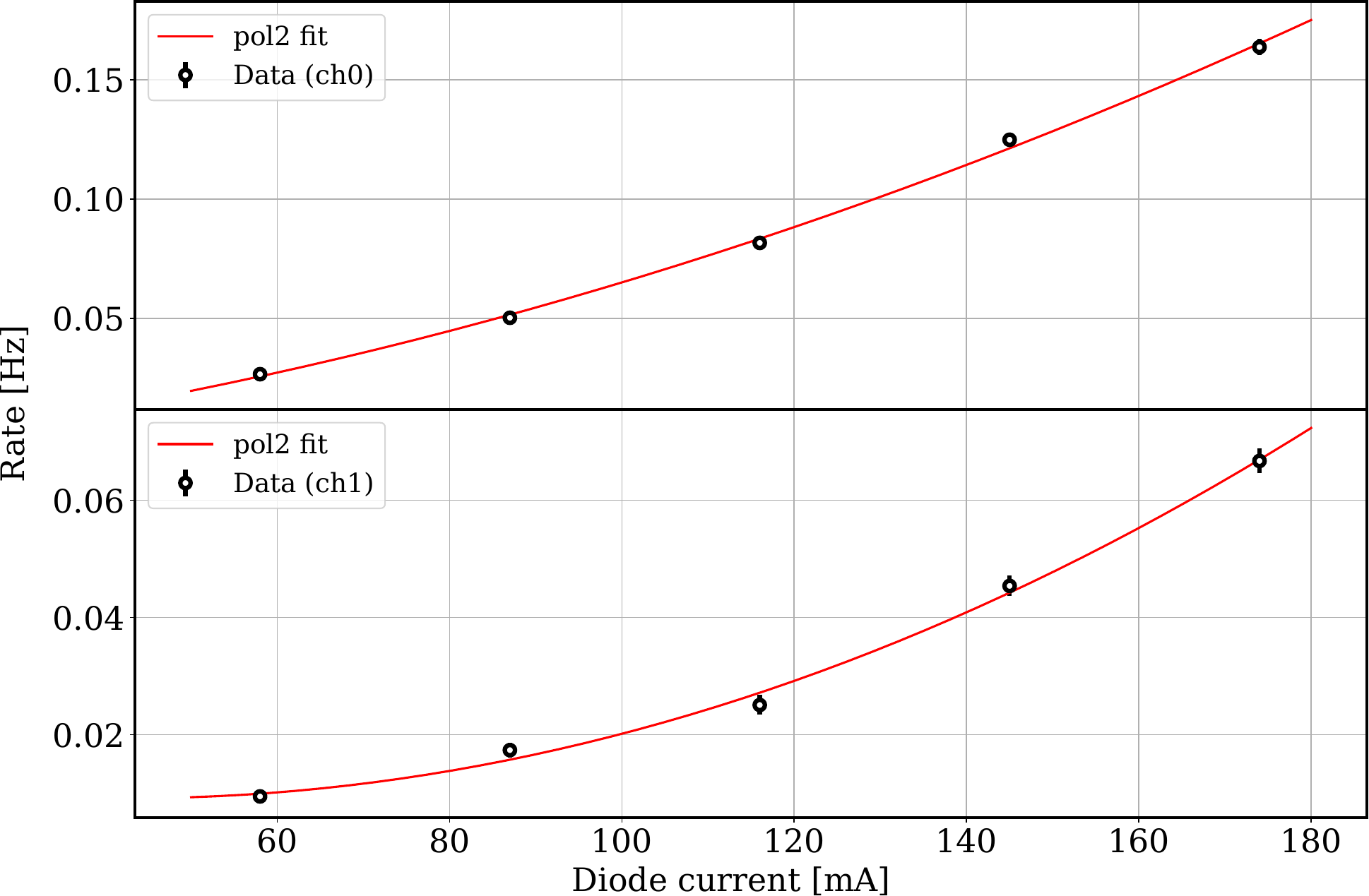}}
\caption{Measured rates for two test channels at five different LED current values. Data were fitted using quadratic (red line) polynomial.}
\label{fig:analysis3}
\end{figure}

We then characterized the signal dependencies. The voltage difference between the Al foils and the TES was set to -300 V. In principle, the event rate is expected to scale with the LED intensity while remaining independent of the voltage applied to the aluminum foils. Conversely, an increase in the applied voltage should result in a corresponding increase in signal amplitude.
The LED current was varied to count the number of detected particles in two of the most active and best-performing channels (0 and 1). As shown in Figure~\ref{fig:arrayholmes}, Channel 0 (1) refers to the thirteenth (twelfth) channel from the left in the second (first) row. Figure~\ref{fig:analysis3} shows the rate as a function of current. Measurements were performed at LED currents of 58, 87, 116, 145, and 174~mA, while maintaining stable cryostat temperatures to keep the TESs at their nominal operating point. Measured rates ranged from 25 to 160 mHz in channel 0 and 10 to 65 mHz in channel 1. This discrepancy is attributed to non-optimal fiber alignment, which resulted in non-uniform illumination of the detector array, as well as differing trigger thresholds. Although the absolute count rates differed between the two channels, the rate ratio remained consistent across the measurements. 
As expected for photoelectrons, the count rate increased with the LED current. The LED datasheet indicates a nearly linear dependence of current on the applied voltage in the 60 to 180 mA range. Under this assumption, the electrical power would scale approximately as the square of the current. Consistently, the measured rates are well described by a quadratic dependence on the current.
Since the rate scales with the LED intensity and the signal appears only when a negative voltage is applied, we can confirm that these are photo‑emitted electrons.

\begin{figure}[h!]
\centering
\resizebox{0.65\textwidth}{!}{\includegraphics{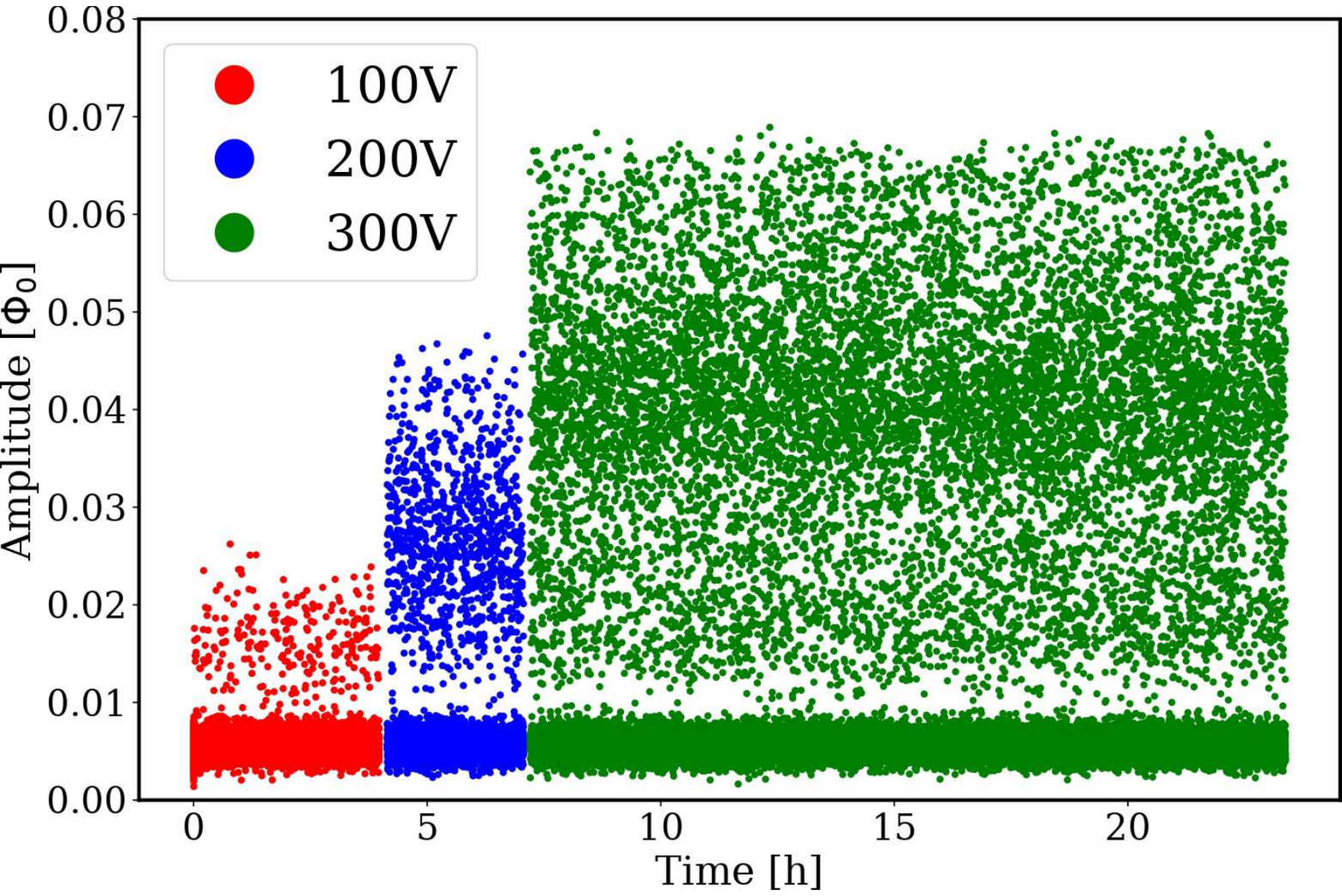}}
\caption{The peak-to-peak amplitude (channel 0) evolution in time of the occurring events in a TES detector according to three different voltage values. The legend shows the absolute value of applied voltage (red 100 V, blue 200 V and green 300 V). Noise contribution dominates at low amplitudes. }
\label{fig:analysis2}
\end{figure}

To investigate the evolution of signal amplitude, measurements were carried out at three distinct negative bias voltages. Figure~\ref{fig:analysis2} displays the peak-to-peak amplitude as a function of time. The datasets correspond to applied potentials of -100 V (red), -200 V (blue), and -300 V (green). Consistent with expectations, increasing the magnitude of the applied potential results in higher energy and a corresponding increase in measured amplitude.

\section{\label{sec:caratterizzazione}Source characterization}
After obtaining evidence consistent with the detected particles being electrons, the next step was to characterize the source, focusing on its beam profile, rate, and efficiency. The room-temperature fiber-coupled LED is driven at 174 mA, while a bias voltage of -300 V is applied to the aluminum foils. A common trigger threshold is set for all channels.

\begin{figure}[h!]
\centering
\resizebox{0.65\textwidth}{!}{\includegraphics{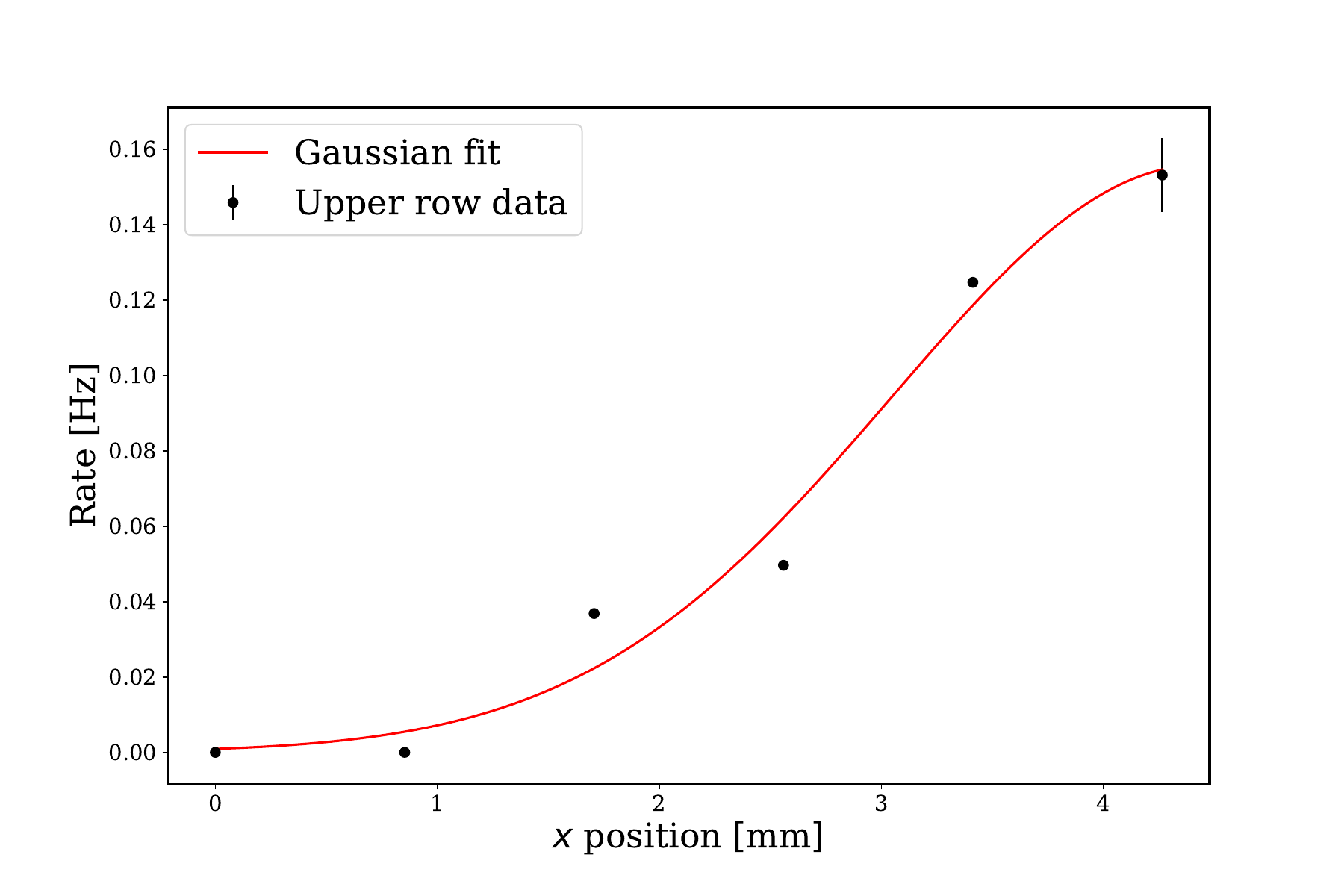}}
\caption{The transverse profile of the electron beam. The black points represent the array upper row rate data and were fitted assuming a normal distribution. The fit suggests a beam profile width of $\sim$ 1.38~mm.}
\label{fig:analysis6}
\end{figure}

A nonzero electron rate is observed only in the channels located in the upper-right region of the array, indicating that the fiber was not perfectly aligned with the detector. This misalignment is attributed to the finite distance between the source and the TES array, which makes the positioning of the fiber with respect to the detector non-trivial. As a result, only the TESs in the upper-right corner of the array are directly exposed to the electrons emitted by the fiber.
To estimate the electron rate, an amplitude cut was applied to the events, since a fraction of the detected signals can be attributed to backscattering or secondary processes, as discussed in more detail in Section~\ref{sec:spectrum}. The threshold was set at 5$\sigma$ above the noise-event distribution. Events exceeding this energy threshold are therefore attributed to electrons directly impinging on the detector. By qualitatively fitting the data from the upper row with a normal distribution, the electron beam profile along the x-axis is roughly extrapolated. The lower row was excluded from the analysis because the second channel from the right exhibited a significantly higher noise level, and for this reason, a higher trigger threshold was used, making its response not directly comparable to that of the others. Therefore, only the upper-row data were used for the profile extrapolation. 
The fit, shown in Figure~\ref{fig:analysis6}, yields a beam profile width of $1.38 \pm 0.33$~mm. This result is consistent with a simple simulation in which electrons are emitted from the Al foil with random directions from a spot corresponding to the fiber diameter and with an initial kinetic energy approximately equal to the photon beam energy minus the aluminum work function. The electrons are then accelerated by an electric field perpendicular to the detector plane. The resulting simulated transverse profile has a standard deviation of about 1.1~mm, in good agreement with the measured value. However, this simple simulation does not account for the misalignment of the fiber, the finite interaction of photons or electrons inside the aluminum foils—which may effectively increase the emission area of the electrons—or geometrical acceptance effects.

Assuming a 2D Gaussian profile, the electron rate at the source is inferred to be $\geq 1$~Hz, and, considering the photon rate, the electron production efficiency is estimated to be $\gtrsim 10^{-14}\ e^-/\gamma$.

\begin{figure}[h!]
\centering
\resizebox{0.65\textwidth}{!}{\includegraphics{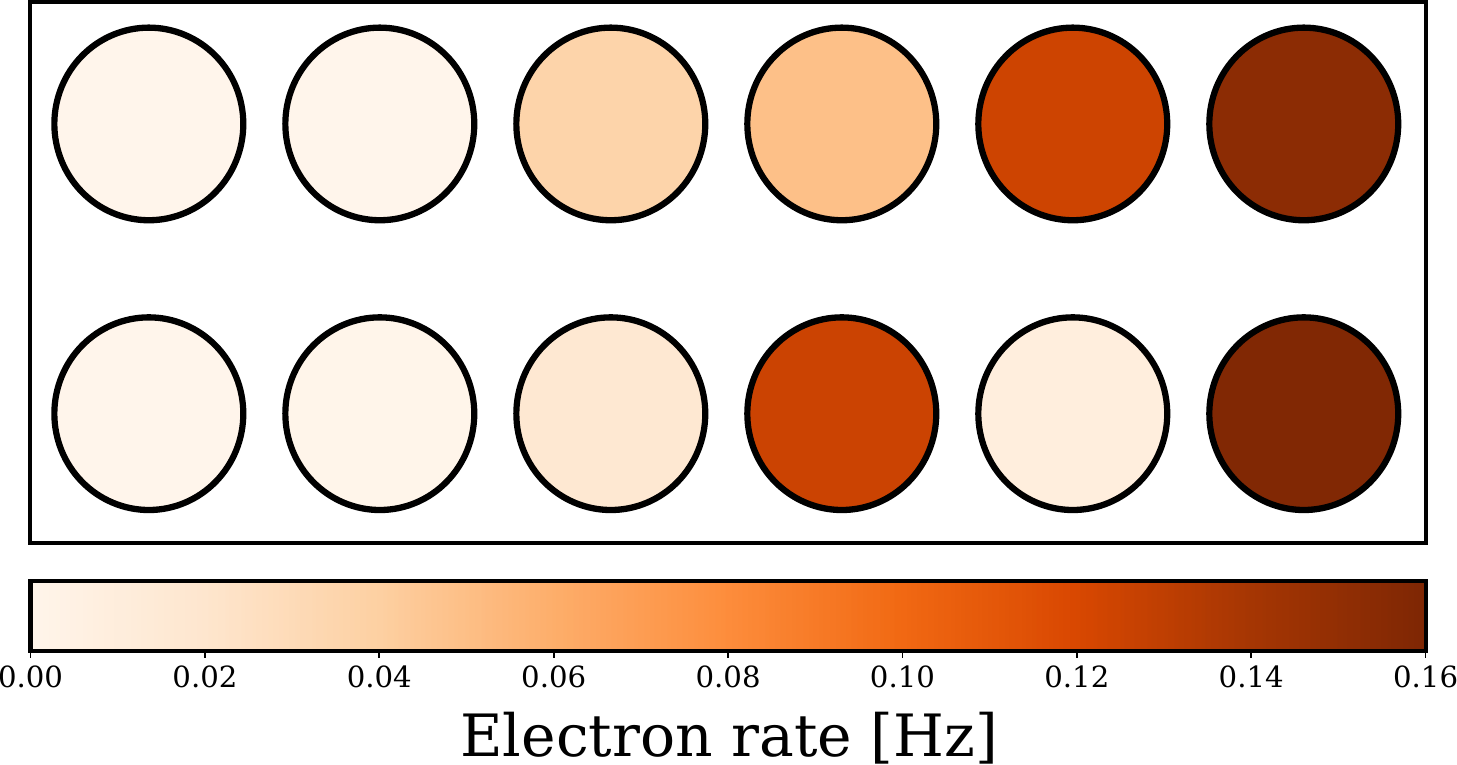}}
\caption{The activity map of our array with a LED current of 174 mA. The fiber orientation aligns with the right section of the array, leading to a significant electron rate in only a few channels. The color scale ranges from white (zero counts) to dark red (0.16~Hz). Channel 0 is the last channel  to the right in the second row, while Channel 1 is the fourth channel in the first row, counting from the left.}
\label{fig:analysis5}
\end{figure}

\section{Detector response to low-energy electrons}\label{sec:spectrum}
It should be emphasized that the primary goal of this work is not to perform precise measurements of very low‑energy electrons, but rather to verify their presence and to characterize the source. It should also be noted that no collimator was available for this measurement, allowing electrons to strike any region of the detector in addition to the absorber and to produce signals with different shapes.

Nevertheless, this section aimed to understand the behavior of impinging low-energy electrons and to investigate the detector response to such electrons. The energy distribution of electrons incident on the gold target was studied through simulations performed with Nebula~\cite{nebula}. Two distinct components emerge from the simulation: a fully elastic contribution, which does not deposit any signal in the detector, and a low–energy component comprising secondary and inelastically backscattered electrons, which are expected to have energies $\lesssim$50 eV, regardless of the initial beam energy. If emitted at sufficiently small angles, with respect to the detector axis, they are expected to be redirected back into the detector by the electric field or into the detector surroundings, contributing to a low–energy background.

Data were recorded using a simple differential trigger. The threshold was set to the smallest value that avoided excessive triggering noise rate.
The sampling time is 4$\mu$s. 
To determine the electron energy distribution, pulses were recorded within acquisition windows including a pre--trigger interval of approximately 10\% for baseline estimation~\cite{Borghesi:2022juf},\cite{Gatti1986},\cite{borghesi_first_2022}. The total window duration was selected to ensure that the microcalorimeter signal fully returned to the baseline.

\begin{figure}[h!]
\centering
\resizebox{0.8\textwidth}{!}{\includegraphics{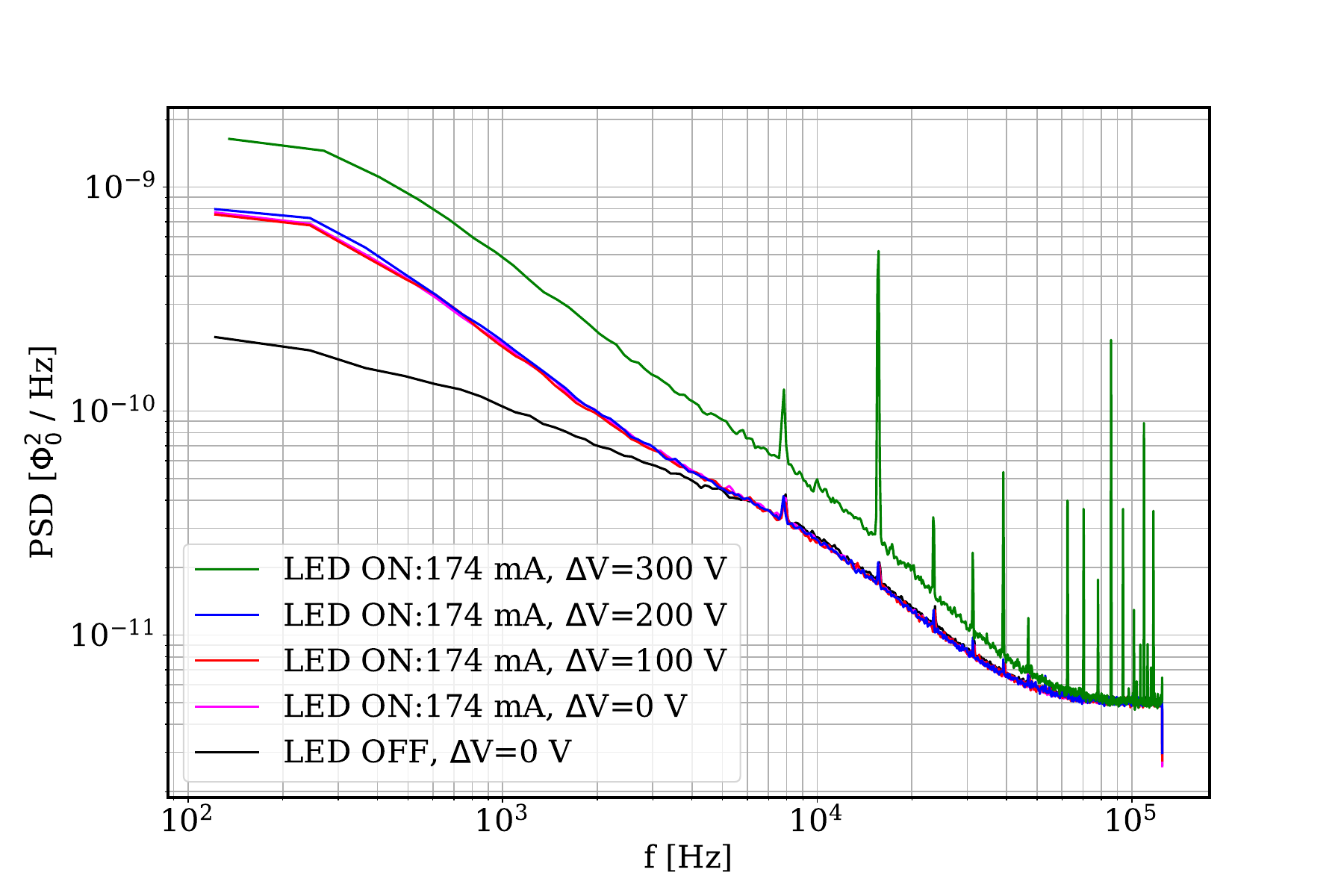}}
\caption{Power Spectral Density with LED off (black), LED ON with $\Delta V=0$ V (magenta), LED ON with
$\Delta V =100$ V (red), LED ON with
$\Delta V =200$ V (blue), and LED ON with
$\Delta V =300$ V (green) for channel 0.}
\label{fig:noise}
\end{figure}

Figure \ref{fig:noise} shows the Power Spectral Density (PSD) for five configurations: LED off, LED on with $\Delta V=0$, and LED on with $\Delta V>0$ (100-200-300 V). The spectral density was computed by averaging a number of spectra obtained by computing a FFT on demodulated records~\cite{bennett2025impactembedded163hoperformance},\cite{Becker_2019}. When comparing the LED‑off condition with the LED‑on case at zero applied potential, the low frequency noise increases by nearly a factor of four. This increase is attributed to photons transmitted through the photocathode. A further rise is observed once an accelerating potential of 300 V is applied. As reported in~\cite{CIMINO2020146876}, the Secondary Electron Yield (SEY) strongly depends on the energy of the incident electrons, reaching a maximum in the few‑hundred‑eV range. This results in an enhanced electron flux at the sensor, which appears as an additional contribution to the measured noise.

Given the lack of collimation and the perturbation due to the secondary electron background described above, in estimating the deposited energy we adopted a method that does not rely on the pulse shape---since events with the same deposited energy may exhibit different morphologies---but instead uses the pulse integral, or \textit{Joule energy}~\cite{Fowler2018}. Additionally, the pulse tail was truncated to avoid integrating contributions from very low--energy events which perturb the baseline. Given that the typical pulse decay time is approximately 300~$\mu$s, the integral was evaluated over this time window.
In the initial stage of data reduction, selection criteria were applied to retain only high--quality events, rejecting signals with very low deposited energy. Then, pulse amplitudes were extracted using the \textit{Joule energy}~\cite{Fowler2018}.
The \textit{Joule energy} of a TES pulse is defined as the time integral of the reduction in Joule power dissipated in the TES relative to its quiescent state~\cite{Fowler2018}. In this
description, the \textit{Joule energy} can be written as a linear combination of the time integrals of the signal $s(t)$ and its square, with coefficients determined by the electrical parameters of the bias circuit, extracted from the TES IV curves:
\begin{equation}
    E_{J}
= R_{\mathrm{sh}}(I_{\mathrm{bias}} - 2 I_q)
  \int_{t_0}^{t_f} s(t)\, dt
+ R_{\mathrm{sh}}
  \int_{t_0}^{t_f} s^2(t)\, dt.
\end{equation}

Here $s(t) = I_q - I(t)$ is the positive-going signal, $I_q$ denotes the quiescent TES current and $R_{sh}$ is the shunt resistor value.

After estimating the \textit{Joule energy}, the spectra were fitted with the following function, introduced in~\cite{BLAND19981225}:
\begin{equation}\label{eq:1}
    \mathcal{S}(E_{J}) = \bigl(\mathcal{S}_{\mathrm{ele}}(E_{J})\cdot N_{\mathrm{ev}}\bigr) * \mathcal{R} + b ,
\end{equation}
where $ N_{\mathrm{ev}}$ is the number of events, $b$ is the flat background, and $\mathcal{R}$ is a Gaussian function centered at zero that models the effective detector response, which is convolved with  
\begin{equation}\label{eq:2}
    \mathcal{S}_{\mathrm{ele}}(E_{J}) = \frac{e^{-E_{J}/\tau}}{\tau}\,\theta(E_{J_0}-E_{J}) ,
\end{equation}
which is the function that describes the behavior of the electrons, where $\theta$ is the Heaviside function, $\tau$ is the tailing parameter, and $E_{J_0}$ corresponds to electrons depositing their full energy in the gold absorber. We find that this functional form provides a good description of the energy distributions obtained from Nebula simulations~\cite{nebula}.

To calibrate the spectra, each of the three datasets (100-200-300 V) was fitted to extract $E_{J_0}$. We perform a Bayesian parameter estimation~\cite{ahrens2026holmes},\cite{alpert2025stringentboundelectronneutrino} using a Poisson likelihood with the model defined in Eqs.~\ref{eq:1} and \ref{eq:2}. The posterior distribution is sampled using a Hamiltonian Markov Chain Monte Carlo algorithm implemented in Stan~\cite{stan}. The fit includes five free parameters: $b$, $N_{\mathrm{ev}}$, $\tau$, FWHM of the effective Gaussian response, and $E_{J_0}$.

\begin{figure}[h!]
\centering
\resizebox{0.8\textwidth}{!}{\includegraphics{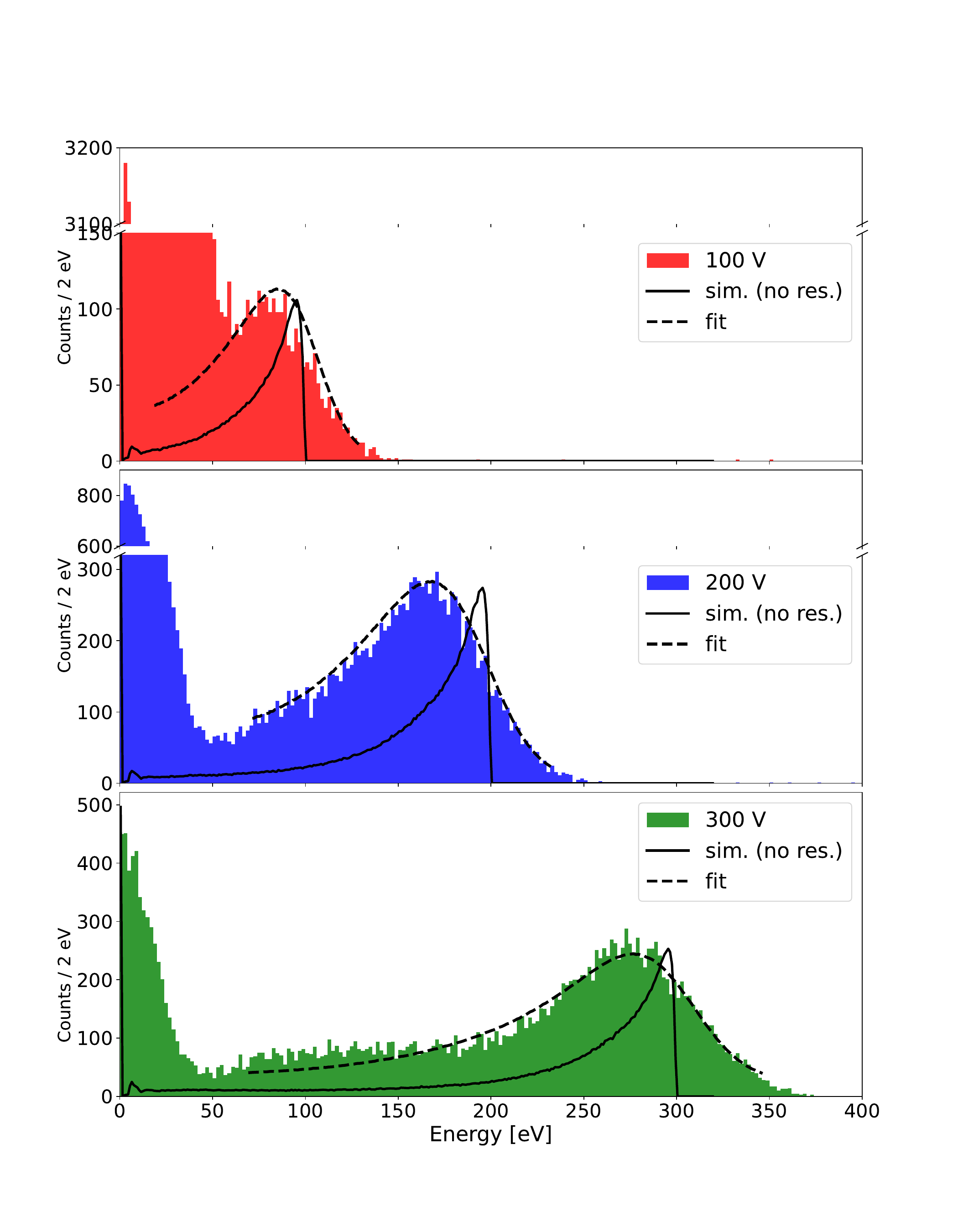}}
\caption{Calibrated energy spectra from channel 0 for different voltages applied between the Al foils and the TESs (red: 100 V, blue: 200 V, green: 300 V). At 100 eV, the signal overlaps with the population of backscattered and secondary electrons, making discrimination more challenging. For this reason, a portion of the 100 eV spectrum has been truncated to maintain signal visibility, while its overall shape remains consistent with that observed in the other channels. The figure also shows the result of the fit procedure (black dashed line) over a narrower energy range, and the results of the simulations performed with Nebula, without accounting for the detector resolution (black solid line).}
\label{fig:analysis7}
\end{figure}

The three spectra are then calibrated using the extracted $E_{J_0}$ as the rescaling parameter. Figure~\ref{fig:analysis7} shows the resulting calibrated energy distributions for channel 0, measured at 100~V (red), 200~V (blue), and 300~V (green).

In addition to the high-energy peak, two other distinct populations can be observed in these distributions. At very low energies ($\lesssim$ 50~eV), the peak is mainly attributed to noise, which includes electrons that deposit only a very small amount of energy in the detector because they undergo elastic or quasi-elastic scattering and are backscattered~\cite{verduin,back2,back}, as well as electrons that produce secondary electrons~\cite{RIDZEL2020146824,sec,bomb,patel_2023}. This background component is also consistent with the simulations. A second low-energy background contribution appears at intermediate energies and corresponds to events with rise times slightly different from those of the main peak, which are associated with interactions occurring in different regions of the detector (e.g., in the 300 V case, in Figure~\ref{fig:analysis7}, they correspond to events in the $\sim$70–170 eV range).
Due to the presence of the different background components, the fitting procedure shown in Figure~\ref{fig:analysis7} (black dashed line) was performed over a reduced energy range, including only the main peak. The results of the Nebula simulations are also shown (black solid line), without considering the convolution with the detector energy resolution. 

To estimate the detector energy resolution for impinging electrons, we used the FWHM values obtained from the fits, in agreement with the energy resolution obtained from the noise peak, which are approximately 40~eV. The measured resolution was found to be consistent across the three investigated configurations. 

The measured energy resolution should not be interpreted as the intrinsic energy resolution of the sensor for two important reasons. First, the broadening of the high-energy peaks is partly due to the fact that electrons do not always deposit their full energy in the absorber; instead, a fraction of the energy may be deposited in other components of the detector. Second, the energy estimator used in this analysis, namely the \textit{Joule energy}, is itself affected by noise~\cite{Fowler2018}. The measured energy resolution is consistent with the expected degradation associated with the use of \textit{Joule energy} instead of the optimum filter, corresponding to a worsening by approximately a factor of five. However, due to the aforementioned effects, the optimum filter estimator could not be adopted in this analysis.

\section{\label{sec:Conclusions}Conclusions}

This work presents the development of a novel compact, switchable electron source designed to operate at cryogenic temperatures. The source is based on a cryogenic electron emitter that exploits photoelectric emission from aluminum. It represents a first step toward the realization of a calibration tool for low-temperature experiments such as HOLMES, where precise calibration is crucial due to the non-linear energy response of TES detectors. Ultimately, the source is intended to generate X-rays beyond the endpoint of the Ho spectrum for calibration purposes.
 
Electrons are produced via photoelectric emission from two 400‑nm‑thick aluminum layers illuminated by an LED inside a dilution refrigerator operating below 100 mK. Our photocathode quantum efficiency is estimated to be $\gtrsim 10^{-14}\ e^-/\gamma$. 
All other observed quantities — such as the electron rate, the beam profile, and the energy distribution — depend primarily on external parameters (LED current, illumination geometry, and acceleration voltage) and are therefore not characteristic of the source itself.
Although not the primary goal of this work, TES detectors developed for the HOLMES experiment successfully detected electrons with energies between 100 eV and 300 eV, demonstrating sensitivity in this low‑energy regime, with energy resolution of about 40 eV. Although a collimator was not available for the present measurements, future studies will make use of one to prevent electrons from striking areas other than the TES.

Future work will focus on modifying the wiring of the photocathode biasing circuit to enable measurements at higher energies, up to the keV scale. Further investigations will also be useful to better understand the processes of electron production and extraction when aluminum is in the superconducting state. In particular, the photon mean free path in aluminum and the electron range at low temperatures are not yet well established, and their influence on the observed emission remains to be investigated. Moreover, it is still unknown how the electron production rate scales with aluminum thickness, with the number of layers, and whether it shows any dependence on the wavelength of the incident light.

Additional steps are also required to develop a viable X-ray calibration source: amplification of the emitted electrons through an electron multiplier, followed by their interaction with an appropriate conversion material to generate X-ray with energy extending up to the endpoint of the $^{163}$Ho spectrum. Several challenges may arise, such as limitations on the maximum LED current imposed by cryogenic operation and background contributions in the final conversion stage from bremsstrahlung, as well as transmitted and secondary electrons. These aspects will be addressed in future studies.
\appendix
\section*{Acknowledgements}\label{acknowledgements}
This work was supported by TES4e, a PRIN 2022, funded by the European Union - Next Generation EU, Mission 4 Component 1, CUP H53C24000870006. The HOLMES experiment is supported by the Istituto Nazionale di Fisica Nucleare (INFN) and by the European Research Council under the European Union’s Seventh Framework Programme (FP7/2007–2013), ERC Grant Agreement no. 340321. Bennett, Mates, Schmidt, Swetz, and Ullom were supported by NIST internal resources. Certain commercial equipment, instruments, or materials, commercial or non-commercial, are identified in this paper in order to specify the experimental procedure adequately. Such identification does not imply recommendation or endorsement of any product or service by NIST, nor does it imply that the materials or equipment identified are necessarily the best available for the purpose.
\section*{Declarations}

\section*{Funding}
This work was supported by TES4e, a PRIN 2022, funded by the European Union – Next Generation EU, Mission 4 Component 1, CUP H53C24000870006. The HOLMES experiment is supported by the Istituto Nazionale di Fisica Nucleare (INFN) and by the European Research Council under the European Union’s Seventh Framework Programme (FP7/2007–2013), ERC Grant Agreement No. 340321. Bennett, Mates, Schmidt, Swetz, and Ullom were supported by NIST internal resources.

\section*{Competing Interests}
The authors declare that they have no competing interests.

\section*{Data Availability Statement}
The datasets acquired during the current study are available from the corresponding author on reasonable request.
\bibliographystyle{elsarticle-num}
\bibliography{refs}

\end{document}